\documentstyle[11pt,paspconf,psfig]{article}
\begin{document}

\title{Dwarf Galaxies in Clusters: the Effects of a Violent Environment} 

\author{Omar L\'opez-Cruz}

\affil{INAOE-Tonantzintla, Luis Enrique Erro No.1, Tonantzintla,
72840 Puebla. M\'exico}

\author{H.K.C. Yee}

\affil{University of Toronto, Department of Astronomy, 60 St. George St.,
Toronto ON. M5S 3H8, Canada}


\section{Introduction}
New observations suggest that dwarf galaxies pervade the universe, for
they have been encountered in large numbers in the field (Marzke et
al. 1998), groups (C\^ot\'e et al. 1977), and clusters of galaxies
(e.g. Trentham 1998). Some early studies encountered an over-abundance
of dwarfs in clusters (e.g. Sandage et al. 1985) in the nearby
universe. Studies at higher redshifts also suggested large numbers of
dwarfs in clusters. The faint-end slope of the luminosity function
(LF) has indicated the overabundance of dwarf galaxies, having steep
values of $-2.0\,\leq\,\alpha\,\leq\,-1.4$. Even a pattern of
universality in the steep slope has been suggested (Trentham 1998).

In the recent studies of clusters' LFs, a factor that has been often
overlooked is the effect of the environment.  For instance, the
characteristics of cD galaxies have always suggested that their
formation is due to processes pertaining to clusters, since no cD
galaxy have been found in the field (Kormendy \& Djorgovsky 1989).  In
general cD clusters (Root-Sastry cD; Boutz-Morgan class I, I-II) are
rich, massive, and dynamically evolved. 
In contrast non-cD clusters are often irregular, poorer, and with less
hot gas in the intracluster medium (ICM). These cluster morphological
and dynamical properties defined mainly by the giant galaxies should
also be reflected in the dwarf population. L\'opez-Cruz \& Yee (1995)
encountered a class of clusters that they termed {\sf flat-LF}
clusters that challenged the idea that dwarf galaxies were very
abundant in clusters' environments. Later L\'opez-Cruz et al. (1997)
reported seven {\sf flat-LF} clusters in a survey of 45 and suggested
that the luminous halos of cD galaxies and a large fraction of the
intracluster gas was formed by the disruption of dwarf galaxies.  The
results of L\'opez-Cruz et al. pointed out that the effects of the
environment were detectable in the dwarf population easier than in the
giant population.

The idea that the cluster environment is hostile to galaxies is
strengthened by new high resolution numerical simulations showing that
the cluster environment is able to strip gas and the outer halos of
giant galaxies, even induce morphological changes (e.g., Moore et
al. 1999).  Dark matter dominated dwarfs should be susceptible to all
the dynamical effects induced by the environment. In some cases the
effects could be devastating, causing the disruption of a large number
of dwarf galaxies, and hence the flat faint-end slopes. A plausible
mechanism is that in rich clusters a dwarf galaxy could suffer
multiple encounters with giant galaxies. Therefore, dwarf galaxies
could be affected by tidal disruption and orbital heating.
Alternatively, the lack of dwarf galaxies in rich environments has
been interpreted in terms of the density-morphology relation: dwarf
galaxies prefer low density environments (Phillipps et al. 1998). We
argue that such an explanation cannot account for the relationship
between the cD halo luminosity and the gas mass in the ICM, whereas
the dwarf disruption scenario addresses this naturally (L\'opez-Cruz
et al. 1997).

\section{Observations}
The observations come from the Low-Redshift Optical Survey (LOCOS)
(see L\'opez-Cruz 1997, Ph.D. thesis).  The LOCOS is a comprehensive
multicolor photometric survey that includes a sample of 45 Abell
clusters with $z$ between $\sim 0.4$ to $0.15$.  The observations were
carried out in Kron-Cousins $I$, $R$ and $B$ bands at
KPNO\footnote{KPNO, NOAO, is operated by the Association of
Universities for Research in Astronomy, Inc. (AURA) under cooperative
agreement with the NSF.}  with the 0.9m telescope and the T2KA CCD
($2048\,\times\,2048$ pixels). The field covered by this combination
is $23\arcmin\,\times\,23.2\arcmin$ with a scale of $0.68\arcsec/{\rm
pixel}$. The integration times varied from 900 to 2500 seconds,
depending on the filter and the redshift of the cluster. The
photometric calibrations were done using stars from Landolt (1992)
compilation. Control fields are also an integral part of this survey,
we observed 5 control fields (in $R$ and $B$) chosen at random
positions in the sky of $5\deg$ away from a cluster observation.  Data
preprocessing was done with IRAF, while the object finding,
star/galaxy classification, photometry, and the generation of
catalogues was done with {\sf PPP} (Yee et al. 1996).

\section{Results}
L\'opez-Cruz \& Yee (1995) pointed out that rich cD clusters had the
tendency towards flatter LFs. More examples were presented in
L\'opez-Cruz et al.  (1997) and Driver et al. (1998). The Schechter
function has been adopted almost universally to describe the behavior
of the LFs. However, this approach has a number of problems. For
example, in many cases there are not enough degrees of freedom. Hence
to avoid degenaracy in the fits, it is necessary to restrict the
parameter space, thereby resulting in an interdependence of the
parameters. This makes the process model dependent.  One can get
around such problems using the data themselves and introducing a
non-parametric dwarf-to-giant ratio (D/G) defined by summing the
counts in the tabulated (in 0.5 mag bins) LFs in the following manner:
the counts in the interval $-20 < {\rm M}_{R} \le -17$ and the counts
of galaxies brighter than $-20$ are summed to give the number of dwarf
and giant galaxies, respectively; i.e.,

\begin{equation}
{\rm D/G}= \frac{N( -20 < {\rm M}_{R} \le-17)}{N({\rm M}_{R} \le -20)}.
\end{equation}

In order to include the largest possible number of clusters to
generate the D/Gs we have included clusters that were complete to
${\rm M}_{R}=-17.25$ by adding an extra bin centered on ${\rm
M}_{R}=-17.0$ with the same counts as in the bin centered on ${\rm
M}_{R}=-17.5$. Figure 1 shows the behavior of D/G versus the richness
parameter $B_{gc}$ (Yee \& L\'opez-Cruz 1999). A clear trend towards
lower D/Gs is seen as richness increases, a simple fit shows that
${\rm D/G}\,\approx\,909B_{gc}^{-1}+2$. It is also seen that cD
clusters tend towards lower D/G than non-cD clusters.  If the dwarfs
are contributing to the gas in the cluster, then the D/G should
correlate with the gas mass ($M_{gas}$; Jones \& Forman 1999) of the
ICM. Such a correlation is seen in Figure 2.

\begin{figure}[t]
\centerline{\hbox{
\psfig{figure=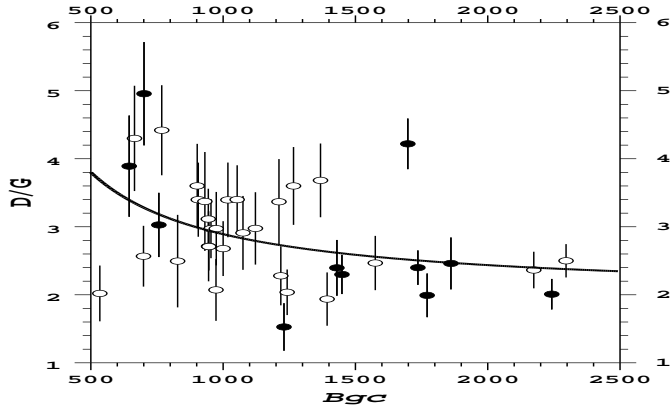,height=7 cm,width=12 cm}
}}
\caption{The variation of D/G with richness ($B_{gc}$). The filled
dots are cD clusters and the solid line shows the fit D/G$\approx
909B_{gc}^{-1}+ 2$. 
\label{figure1.eps}}
\end{figure}

\section{Discussion and Conclusion}
 We have disregarded low surface brightness (LSB) galaxies in our
discussion, although they have been detected in large numbers in poor
clusters such as Virgo (Impey et al. 1998). We claim that our results
are independent of our efficiency to detect (LSB) galaxies since we
have included fairly bright objects. Our 100 \% completeness limit is
0.9 mag brighter than the $5\sigma$ stellar detection limit. Moreover,
LSB galaxies are unlikely to survive in clusters (Moore et al.  1999).
There are two main conclusions drawn from the present study. First
that there is a range of D/G that seems to be correlated with the
richness of the cluster. This is also indicative of a wide range in
faint-end slopes in the LFs of cluster galaxies. If this trend of
decreasing number of dwarfs with richness is extended to very poor
regions, then we propose a ${\rm D/G}\sim 15$ for the field
($B_{gg}\approx 68$). Second, the relationship between the gas mass of
the ICM and the D/G supports the dwarf disruption scenario for the
formation of the luminous halo of cD galaxies and the origin of the
ICM. The alternative explanation of the D/G behavior in terms of dwarf
population density relation cannot account for the relationships
between the ICM gas mass and the luminosity of the cD halo
(L\'opez-Cruz et al. 1997). It also cannot account for the relationship
with the D/G presented above.

\begin{figure}[t]
\centerline{\hbox{
\psfig{figure=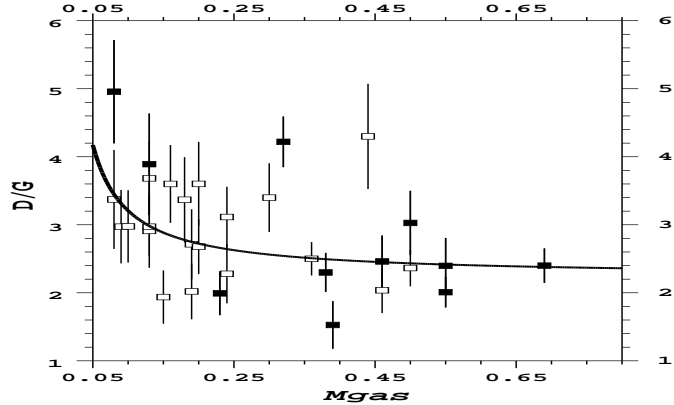,height=7 cm,width=12 cm}
}}
\caption{The variation of D/G with the gas mass
$M_{gas}[10^{14}M_{\sun}$]. The filled squares are cD clusters and the
solid line is the fit D/G$\approx 1\times 10^{14}M_{gas}^{-1}+ 2$.
If disrupted dwarfs have contributed to the gas in the ICM, then this
correlation is expected.
\label{figure2.eps}}
\end{figure}

\end{document}